# Fast Switching Ferroelectric Materials for Accelerator Applications

A. Kanareykin[1], E. Nenasheva[2], V. Yakovlev[3], A. Dedyk[4], S. Karmanenko[4], A. Kozyrev[4], V. Osadchy[4], D. Kosmin[4], P. Schoessow[1], A. Semenov[4].

[1] Euclid Techlabs LLC, Solon, OH 44139, USA
[2] Ceramics Co. Ltd, St. Petersburg 194223, Russia
[3] Omega-P Inc., New Haven, CT 06511, USA
[4] St. Petersburg Electrical Engineering University, St. Petersburg 197376, Russia

**Abstract**. Fast switching (< 10 nsec) measurement results on the recently developed BST(M) (barium strontium titanium oxide composition with magnesium-based additions) ferroelectric materials are presented. These materials can be used as the basis for new advanced technology components suitable for high-gradient accelerators. A ferroelectric ceramic has an electric field-dependent dielectric permittivity that can be altered by applying a bias voltage. Ferroelectric materials offer significant benefits for linear collider applications, in particular, for switching and control elements where a very short response time of <10 nsec is required. The measurement results presented here show that the new BST(M) ceramic exhibits a high tunability factor: a bias field of 40-50 kV/cm reduces the permittivity by a factor of 1.3-1.5. The recently developed technology of gold biasing contact deposition on large diameter (110 cm) thin wall ferroelectric rings allowed ~few nsec switching times in witness sample experiments. The ferroelectric rings can be used at high pulsed power (tens of megawatts) for X-band components as well as at high average power in the range of a few kilowatts for the L-band phase-shifter, under development for optimization of the ILC rf coupling. Accelerator applications include fast active X-band and Ka-band high-power ferroelectric switches, high-power X-band and L-band phase shifters, and tunable dielectric-loaded accelerating structures.

## INTRODUCTION

A ferroelectric ceramic is a material with an electric-field-dependent dielectric permittivity that can be altered very rapidly by an applied bias voltage pulse. Typical representative ferroelectric materials are $(Ba,Sr)TiO_3$ or a $BaTiO_3$ - $SrTiO_3$ solid solution (BST). The BST material can be synthesized in polycrystalline, ceramic layer, and bulk forms [1-3]. Ferroelectrics have unique intrinsic properties that make them extremely attractive for high-energy accelerator applications. The response time is $~10^{-11}$ sec for crystalline and $~10^{-10}$ sec for ceramic compounds [1]. Unlike semiconductors and plasma devices, ferroelectrics allow control of their dielectric properties in two directions (varying the dielectric parameters and restoring the initial phase) using a single external control pulse, offering unique capabilities for high-power switching device design. High dielectric breakdown strength, low gas permeability and simplicity of mechanical treatment make ferroelectric ceramics promising candidates for the loading material in accelerator tuning and switching devices [4-6].

Fast switching low-loss ferroelectric materials [7-8] offer significant advantages for rf adjustment and tuning in both advanced and more conventional accelerators, in particular for applications involving switching and high power control [9-12]. Ferroelectric-based devices that have been recently proposed include: (1) ultra-fast electrically-controlled phase shifters for amplitude and phase control of superconducting and room temperature accelerators including ILC and ERLs [9]; (2) high-power switches for pulse compression systems of stand-alone sources for high–gradient tests of future CLIC-like linear colliders [10,11]; (3) tunable dielectric loaded accelerating (DLA) structures [13-14].

One application of this technology is relevant to the International Linear Collider (ILC) project [9]. Based on SC accelerating structures, the ILC will accelerate beams of electrons and positrons to 0.5 TeV. The collider will require thousands of SC RF cavities. In the current ILC design about twenty accelerating structures are fed from a single RF source. This plan for RF power distribution requires special devices to be developed that provide individual phase tuning and matching for each accelerating structure. The present ILC project design calls for these devices to be mechanical three-stab tuners. However, by employing ferroelectric ceramics with low losses [6,8] electrically-controlled tuners with switching times of ~few microseconds [7,8] can be designed. Such a tuner allows adjustment of the coupling of the accelerating structure over timescales much smaller than the duration of the RF pulse. In turn this capability allows reduction of cryogenic losses at the structure and eventually significant reductions of the entire energy consumption and operating cost of the proposed collider scheme [9]. This type of fast tuner can also be employed in many other SC and conventional accelerators for amplitude and phase control optimization [9-12], for example in ERL systems.

This new ferroelectric composition has direct applications to development of tunable dielectric loaded accelerating (DLA) structures [13]. Frequency tuning (or phase velocity adjustment) will allow the elimination of any frequency shift caused by machining tolerances of the dielectric dimensions, thermal expansion of the structure, or dielectric constant heterogeneity [13-14]. The method used to vary the frequency of a DLA structure consists of incorporating a thin layer of a ferroelectric material backing a layer of conventional ceramic. A DC bias voltage is used to vary the permittivity of the ferroelectric layer and thus tune the overall frequency of the DLA structure [13-14].

The properties and parameters achieved for some of these low-loss BST-based ferroelectric ceramics were published previously [4-6,8]. In this paper we present new results on the development of BST(M) ferroelectric material. Special attention will be paid to the improvements in the time response of the material. The technology of metallization of the ferroelectric ceramic that is critical to achieve time responses of < 10 nsec is discussed. The experimental techniques used for the time response measurements are also presented.

This newly developed ferroelectric technology has been experimentally validated at the Argonne Wakefield Accelerator in a tunable DLA structure experiment [14] and is under further development at Omega-P Inc./Yale University for the demonstration of high power microwave switches and fast ferroelectric phase-shifters at both L-band for ILC and at Ka-band for high accelerating gradient research [6,8,9-12].

## FAST SWITCHING AND LOW LOSS BST FERROELECTRIC DEVELOPMENT

**Ferroelectric properties required for accelerator component development**. Ferroelectrics have an **E**-field-dependent dielectric permittivity $\varepsilon$ that can be very rapidly altered by application of a bias voltage pulse. The switching time in most instances is limited by the response time of the external circuits that generate and transmit the high-voltage pulse, and can therefore be in the nanosecond range. Ferroelectric materials should have the following properties in order to be used in high-power rf switches for linear collider applications and tunable DLA structures [6,8]: (1) the dielectric constant should not exceed 300-500 to avoid problems in the switch design caused by interference from high-order modes and extra wall losses; (2) the dielectric constant should be variable by 15-20% to provide the required switching and tuning properties; (3) bias electric fields required to adjust the permittivity within this range should be no larger than a few 10's of kV/cm; (4) the loss tangent should be in the range of $5\times10^{-3}$ or lower at 11 GHz to allow switch operation at high average power in a collider having a repetition rate of 120-180 Hz. DLA structures also require loss factors in the range of $5\times10^{-3}$ at 11.4 GHz.

Along with its electrical properties, the material must have sufficient mechanical strength to permit the manufacturing of complex and massive objects of relatively large volumes that can be machined to high mechanical tolerances. At the same time, the technology of the tunable material fabrication must provide extremely high homogeneity of the dielectric constant and other microwave parameters throughout the volume of the finished ferroelectric elements [8].

Modern bulk ferroelectrics, such as barium strontium titanate ($Ba_xSr_{1-x}TiO_3$, or BST with additives, dielectric constant ~ 500), have rather high electric breakdown strengths (100-200 kV/cm) and do not require too large a bias electric field (~20-50 kV/cm) to effect a significant variation (20-30%) of the dielectric constant ε [3,4-6]. The loss tangent for commercially-available samples of these materials is ~$10^{-2}$ at 10 GHz [2,3]. Euclid Techlabs LLC has recently developed and tested a modified bulk ferroelectric based on a composition of BST ceramics and magnesium-based additives. The parameters of the newly synthesized composition satisfy requirements (1)-(3) listed above [6,8].

**X-band Low Loss Ferroelectric Development**

We recently developed and tested a bulk ferroelectric composed of BST ceramics and magnesium oxides, that exhibits a permittivity ε = 500, dielectric loss factor in the range of $5\times10$-3 and tunability factor of 1.20-1.30 with a 40-50 kV/cm bias field applied. [4-6, 8].

Barium strontium titanate solid solutions $(Ba_xSr_{1-x})TiO_3$ (x = 0.45 – 0.55) were synthesized by ceramic processing from titanium dioxide ($TiO_2$) and strontium and barium carbonates ($SrCO_3$, $BaCO_3$) or presynthesized barium and or strontium titanates ($BaTiO_3$, $SrTiO_3$). The initial materials were treated mechanically by mixing them in a vibration mill for three hours to a particle size of 1 μm before sintering. The sintering temperatures of the ceramic samples were in the range of 1380-1540 °C.

In the papers [4,6], two kinds of ferroelectric compositions based on $(Ba_xSr_{1-x})TiO_3$ (BST) solid solutions with additive of Mg-based compounds (BST(M)-1 and BST(M)-2) have been presented. The ratio of barium and strontium components was in the range x = 0.4-0.5 with a content of up to 20% wt. of magnesium-based compounds. Ceramic samples developed out of these ferroelectric compositions demonstrated the dielectric constant in the range of ε = 400—500. Loss tangents of the samples at 3-11 GHz were found in the range $(4.1-6.8)\times10^{-3}$ for BST(M)-1 and $(2.1-4.2)\times10^{-3}$ for BST(M)-2 [4-5].

The tunability factor (Δε /ε) at low frequency (1 MHz) on samples made of ferroelectric ceramics was measured to be 11% at 2.4 V/μm for BST(M)-1 and 6% at 2.0 V/μm for BST(M)-2. Tunability measurements at microwave frequencies (10 GHz) of the BST(M)-2 samples obtained a tunability factor of ~9.5% at 2.8 V/μm bias field [4-6]. On average the tunability factor is in the 1.15 – 1.20 range at 50 kV/cm for this type of ferroelectric ceramic.

**Ferroelectric-Based Accelerator Component Fabrication**

Initial samples of BST(M)-type ceramic materials studied in this work are powders with particle size ~ 1 μm based on solid solutions of barium/strontium titanates with magnesia-based additives. The composition of these powders as well as their processing technology provides the option of using various methods of forming half-finished products that are widely used in ceramic technology. These are primarily methods of hydraulic and isostatic pressing used for ceramic preforms as cylinders (pipes) of various lengths and cross-sections.

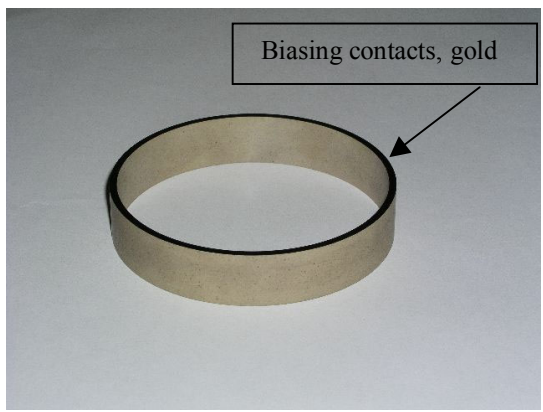

**FIGURE 1.** Prototype BST(M) ferroelectric ring sample. The ring diameter is 110 mm, thickness is 2.8 mm. Note the 2 μm thick gold biasing contact deposited on the ring edge.

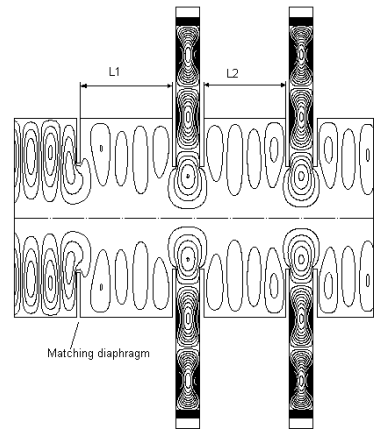

**FIGURE 2.** Conceptual layout of the switch. The operating mode of the cavities is $TE_{031}$.

We produced the basic shapes from ceramic powders similar in composition and processing to those used as dielectric waveguides for DLA applications [13-14]**.** This technology provides uniform hydraulic pressing at the horizontal plane of the ring. At the same time, to fabricate a ferroelectric tube with diameter ~ 100-120 mm made of small grain powders we used a *combination* of *hydraulic* and *isostatic* pressing. This combined technology ensures a highly homogeneous compression of the material along the total ring length to provide uniform dielectric properties along the length of the element.

A specially developed technology of double layer magnetron metal deposition has been used to apply the gold contacts for the bias voltage. This technology allows deposition of a 1-2 μm thick gold layer on a sub-layer at the ferroelectric surface providing an adhesion satisfying the mechanical requirements of the device. The same kind of metallization in turn enables the fast < 10 nsec switching of the ferroelectric. The electrical measurement results are presented in next section.

Fig. 1 shows one such large diameter ferroelectric ring (~110 mm) with the deposited gold biasing contacts. Rings 108 mm in diameter, 20 mm in length and 2.8 mm in thickness have been fabricated for the active ferroelectric RF switch design presented in Fig.2.

**Enhanced Tunability Fast Switching Ferroelectric Development** In this paper we present the development and measurement results for the new BST(M) type ferroelectric materials with the enhanced tunability factors.

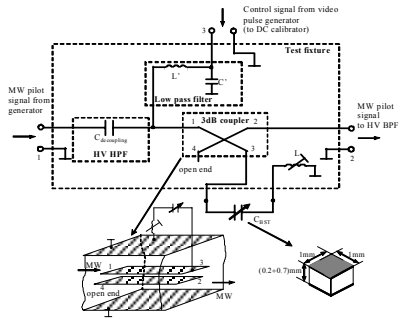 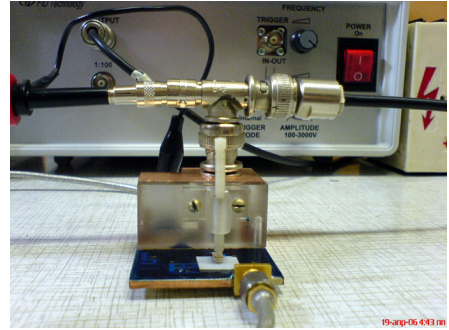

**FIGURE 3.** Experimental setup for measurements of the time response of the ferroelectric samples under HV bias.

**FIGURE 4.** Photo of the test fixture used in the experimental setup for measurements of time response.

The basic parameters of the nonlinear compositions with increased tunability and relatively low microwave loss factor based on the BST ferroelectric solid solutions have been presented in Refs. [1, 3, 4, 5]. Following Ref. [5], it should be noted that the basic principle of the development this kind of composition is the forming of ferroelectric phases with Curie temperatures < 200°K. The tunability factor of the BST-based ferroelectric composition can be enhanced by increasing its barium titanate content. The corresponding Curie temperature shift to a higher temperature range along with the increased tunability causes the dielectric constant and microwave loss factor to increase as well. A BST-based composition especially optimized for the high-power accelerator applications has been synthesized. This material with an increased barium titanate content of $x > 0.5$ exhibits a dielectric constant in the range of 400-600 and contains Mg-based additives (more than 20% wt.). The Mg compounds used here are linear ceramic dielectrics, but interact weakly with the main ferroelectric phase of BST ferroelectric and exhibit a relatively low dielectric constant by comparison with the main BST phase.

In next section are presented fast switching studies of the samples of two kinds of ferroelectric compositions based on $(Ba_xSr_{1-x})TiO_3$ solid solutions with additives of various Mg-compounds (BST(M)-3 and BST(M)-4) with the ratio of barium and strontium components in the range $x = 0.55- 0.60$ but with the addition of magnesium compounds from 25% up to 50% wt. The fast switching dielectric response measurements have been done with gold metallized witness BST(M)-3 and BST(M)-4 samples with dimensions of $1.0 \times 1.0 \times 0.7$ mm$^3$ and $1.0 \times 1.0 \times 0.2$ mm$^3$ respectively.

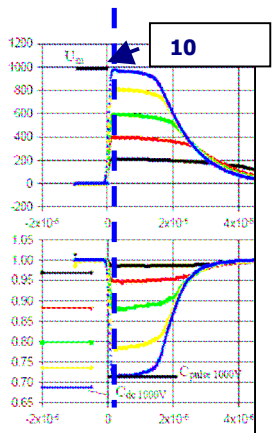 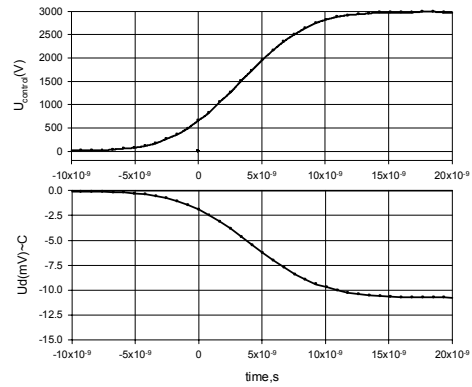

**FIGURE 5.** Ferroelectric time response of a BST(M) sample with the gold biasing contact (and sub-layer), and a 1000 V bias applied. The vertical scale is in μsec.

**FIGURE 6.** Ferroelectric time response measurement results on a nsec scale. BST(M) sample with the gold biasing contact (and sub-layer). Switching time is << 10 nsec.

## Fast Switching RF Measurements

The block-diagram of experimental setup are shown in Fig.3. Fig.4 presents a picture of the test fixture. A microwave signal of f~1.3GHz is fed through a high voltage high pass filter (HV HPF) to the input port 1 of a 3dB- 90° directional coupler. Port 3 of the coupler is loaded by a series resonant circuit including the BST(M) capacitance under test (tunable end). Port 4 is open. The output microwave signal in port 2 is the sum of signals, reflected from the ports 3 and 4. Thus output signal amplitude, or the coupler transmission coefficient S21 depends on the phase shift ($\Delta\varphi$) of the microwave signal reflected from the tunable end. In limiting case of $\Delta\varphi=0$ the output signal falls to zero ($|S_{21}|=0$) and at $\Delta\varphi=180°$ the output signal is maximal ($|S_{21}|=1$.

The microwave detector is sensitive to the envelope of the output signal amplitude. Under high voltage pulses from the biasing HV pulser the capacitance of the metallized ferroelectric is varied, resulting in a corresponding variation of the phase of the reference signal reflected from port 3 that in turn leads to variation of $S_{21}$. A dual channel oscilloscope records the control pulse voltage and the detected reference signal $S_{21}$. A high voltage band-pass filter (HV BPF) is included in the circuit to protect the detector from harmonics of the control voltage from the high voltage pulser. It is found that for the 3 kV biasing pulse effective suppression down to 0.5 mV was successfully provided by the HV BPF. To evaluate the capacitance of the ferroelectric sample under test the experimental setup was calibrated in terms of a normalized capacitance.

Measurement results are presented in Figs. 5 and 6. Note that the test fixture provides no distortion in time scale down to 1 ns for bias pulses applied from the HV pulser to the ferroelectric sample. The shape of the control pulses presented in Fig. 5 (rise time ~4 ns, fall time ~1μs) is defined only by the parameters of the HV pulser. Typical results for the ferroelectric samples are shown on a the μsec scale in Fig. 5 and expanded to the nsec time range in Fig. 6. One can see that the rf signal retraces the bias voltage curve shape demonstrating the short response time of the material. Analysis of the measurement data presented in Fig. 6 shows that the BST(M) ferroelectric with the gold bias contacts provides a switching time considerably less than 10 nsec, much smaller than the switching time needed for our present applications.

## SUMMARY

We have created new BST(M) materials intended for advanced accelerator applications. Our work involved the entire production chain: synthesis of ceramic ferroelectric samples; study of their microstructure; shape and size distributions; dielectric response measurements in the 10-35 GHz frequency range; temperature dependence of the ferroelectric parameters; and electrical properties-- permittivity, loss tangent and tunability. Loss tangents in the range of $(3-4)\times10^{-3}$ at 10 GHz and tunabilities in the range of 25-35 % at 40 - 50 kV/cm biasing fields have been achieved. We have developed large diameter (110 mm) ring samples to be used as RF switching elements for accelerator applications. A double-layer bias contact deposition process has been developed. We have demonstrated switching times in the order of few nsec for BST(M) ferroelectric samples. All these efforts will lead to high power demonstrations of devices incorporating these new ferroelectric materials.

## ACKNOWLEDGMENTS

We would like to thank A.Tagantsev, J.Hirshfield, W.Gai, and J.G.Power for useful discussions. This research was supported by the US Department of Energy, High Energy Physics Division Grant # DE-FG02-04ER83946 and by the Russian Foundation for Basic Research, RFBR Grant # 06-02-16442-a.

## REFERENCES

1. G.A. Smolensky, Ferroelectrics and Related Materials, Academic Press, New York, (1981).
2. O. Vendik, Ferroelectrics at Microwave. O.G.Vendik, Editor, Moscow, Radio, p.44, (1979).
3. A. Tagantsev et al. Journal of Electriceramics, v.11, pp. 5-66, (2003).
4. E.A. Nenasheva, A.D. Kanareykin, N.F. Kartenko, S.F. Karmanenko. Journal of Electroceramics, v.13, (2004).
5. S.F. Karmanenko, A.D. Kanareykin , E.A. Nenasheva , A.I. Dedyk , A.A. Semenov. Integrated Ferroelectrics, v.61, pp.177-181, (2004).
6. A. Kanareykin, E. Nenasheva, S.Karmanenko and V.Yakovlev., AAC-2004, AIP Conference Proceedings v. 737, p.1016-1024, (2004).
7. A.B.Kozyrev et al. Technical Physics Letters, v. 24, pp.19-25, (1998). Private communication (2004).
8. A.Kanareykin, P. Shoessow, E.Nenasheva, S.Karmanenko, A.Dedyk, V.Yakovlev. EPAC-2005, Edinburgh, UK, June 26-30 2006, pp. 3251-3253, (2006).
9. S.Yu. Kazakov, V.P. Yakovlev, J.L. Hirshfield, A.D. Kanareykin, E.A.Nenasheva, "Fast Ferroelectric Tuner For L-Band," these Proceedings.
10. V.P. Yakovlev, O.A. Nezhevenko, and J.L. Hirshfield, PAC2005, Knoxville, May 16-20, 2005, p. 2056-2058, (2005).
11. V.P. Yakovlev et al. RF2003, AIP Conf. Proc., **691**, p. 187 (2003).
12. V.P. Yakovlev, O.A. Nezhevenko , J.L. Hirshfield, AAC-2005, AIP Conference Proceedings v. 737, p.643-650, (2004).
13. A. Kanareykin, W. Gai, J.G.Power, E. Sheinman and A. Altmark. AAC-2002, AIP Conference Proceedings v. 647, p. 565-576 (2002).
14. A.Kanareykin, S.Karmanenko, A.Semenov, E.Nenasheva, P.Schoessow. Proceedings PAC-2005, Knoxville, May 16-20, 2005 pp. 3529-3532. (2005).